\documentclass[journal]{vgtc}

\usepackage{graphicx}
\usepackage{subcaption}

\usepackage{algorithm}
\usepackage{algorithmicx}
\usepackage[noend]{algpseudocode}
\usepackage{tabularx}
\usepackage{makecell}

\usepackage{amsmath}
\usepackage{amssymb}
\usepackage{amsthm}

\usepackage{microtype}
\PassOptionsToPackage{warn}{textcomp}
\usepackage{textcomp}
\usepackage{mathptmx}
\usepackage{times}
\usepackage{cite}

\usepackage{booktabs}

\usepackage[square, sort, numbers]{natbib}

\usepackage{tikz}
\usetikzlibrary{arrows.meta}

\def\legend[#1](#2)(#3)(#4)(#5)(#6){
    \definecolor{c0}{RGB}{255,255,178};
    \filldraw [fill=c0] (0, 8) rectangle (#1, 10);
    \node [right] at (1 + #1, 9) {#2};

    \definecolor{c1}{RGB}{254,204,92};
    \filldraw [fill=c1] (0, 6) rectangle (#1, 8);
    \node [right] at (1 + #1, 7) {#3};

    \definecolor{c2}{RGB}{253,141,60};
    \filldraw [fill=c2] (0, 4) rectangle (#1, 6);
    \node [right] at (1 + #1, 5) {#4};

    \definecolor{c3}{RGB}{240,59,20};
    \filldraw [fill=c3] (0, 2) rectangle (#1, 4);
    \node [right] at (1 + #1, 3) {#5};

    \definecolor{c4}{RGB}{189,0,28};
    \filldraw [fill=c4] (0, 0) rectangle (#1, 2);
    \node [right] at (1 + #1, 1) {#6};
}

\newcommand*\Let[2]{\State #1 $\gets$ #2}
\newcommand*\tab{\hskip\algorithmicindent}
\def\NoNumber#1{{\def\alglinenumber##1{} \\ \hspace{6em} #1}\addtocounter{ALG@line}{-1}}
\algrenewcommand\algorithmicindent{1.0em}%

\usepackage{enumitem}
\setlist[enumerate]{topsep=1pt}
\providecommand{\tightlist}{%
  \setlength{\itemsep}{0pt}\setlength{\parskip}{0pt}}

\onlineid{0}

\vgtccategory{Research}
\vgtcpapertype{please specify}

\title{Equal Area Breaks: A Classification Scheme for Data to Obtain an Evenly-colored Choropleth Map%
\thanks{This work was supported in part by the National Science
Foundation under Grants IIS-12-19023, IIS-13-20791, and IIS-18-16880.}
}

\author{Anis Abboud, John Kastner, and Hanan Samet}
\authorfooter{
\item
 Anis Abboudis with University of Maryland. E-mail: anis@cs.umd.edu.
\item
 John Kastner with University of Maryland. E-mail: kastner@umd.edu.
\item
 Hanan Samet with University of Maryland. E-mail: hjs@cs.umd.edu.
}

\abstract{
An efficient algorithm for computing the choropleth map classification scheme
known as {\em equal area breaks} or {\em geographical quantiles} is introduced.
An equal area breaks classification aims to obtain a coloring for the map such
that the area associated with each of the colors is approximately equal.  This
is meant to be an alternative to an approach that assigns an equal number of
regions with a particular range of property values to each color, called {\em
quantiles}, which could result in the mapped area being dominated by one or a
few colors.  Moreover, it is possible that the other colors are barely
discernible.  This is the case when some regions are much larger than others
(e.g., compare Switzerland with Russia).  A number of algorithms of varying
computational complexity are presented to achieve an equal area assignment to
regions.  They include a pair of greedy algorithms, as well as an optimal
algorithm that is based on dynamic programming.  The classification obtained
from the optimal equal area algorithm is compared with the quantiles and Jenks
natural breaks algorithms and found to be superior from a visual standpoint by
a user study.  Finally, a modified approach is presented which enables users to
vary the extent to which the coloring algorithm satisfies the
conflicting goals of equal area for each color with that of assigning an equal
number of regions to each color.}

\CCScatlist{
  \CCScat{H.5.2}{Information Interfaces and Presentation (e.g., HCI)}{User Interfaces}{map visualization};
  \CCScat{I.1.2}{Symbolic and Algebraic Manipulation}{Algorithms}{sequence partitioning}
}

\vgtcinsertpkg

\begin{document}

\firstsection{Introduction}

\maketitle

Maps are used primarily for navigation and to visualize how measurements vary
across a geographic area.  In this paper we are concerned with the
visualization of spatially-varying ratio measurements (e.g.~population
density).  In particular, we do this using choropleth maps.  In choropleth
maps, region boundaries are predefined but can vary in size and scope.  This is
in contrast with maps such as dasymetric maps whose regions have boundaries
defined by the data values~\cite{slocum_thematic_2009}.

One of the keys to understanding a choropleth map is in understanding the
classification method used to differentiate between the data values.  The data
values can be broken up into classes depending on the ranges of the data values
so that they correspond to intervals that are equal in magnitude, equal in
cardinality or length, as well as equal in area.  The equal area classification
method, which is our focus, is designed to overcome a general drawback of other
classification processes, which, by not taking the area into account, can
result in the map being dominated by one or a few colors, with the other colors
being barely visible.

\begin{figure}
  \centering
  \begin{subfigure}[t]{0.25\columnwidth}
    \includegraphics[width=\columnwidth]{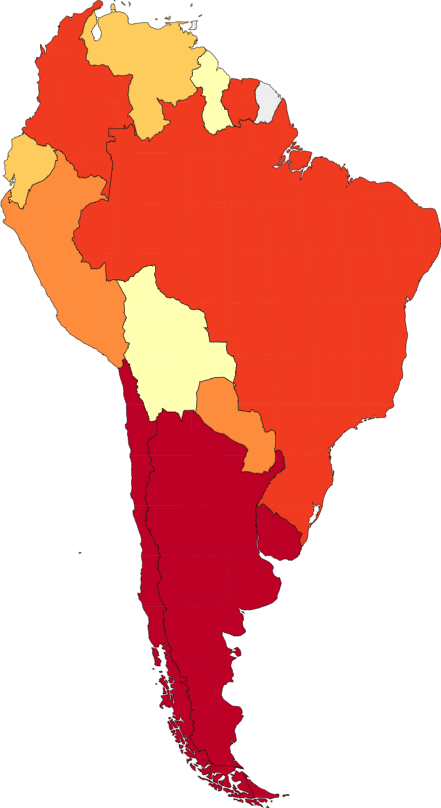}
    \caption{Equal Length}
    \label{fig:south-america-length}
  \end{subfigure}%
  ~
  \begin{subfigure}[t]{0.25\columnwidth}
    \includegraphics[width=\columnwidth]{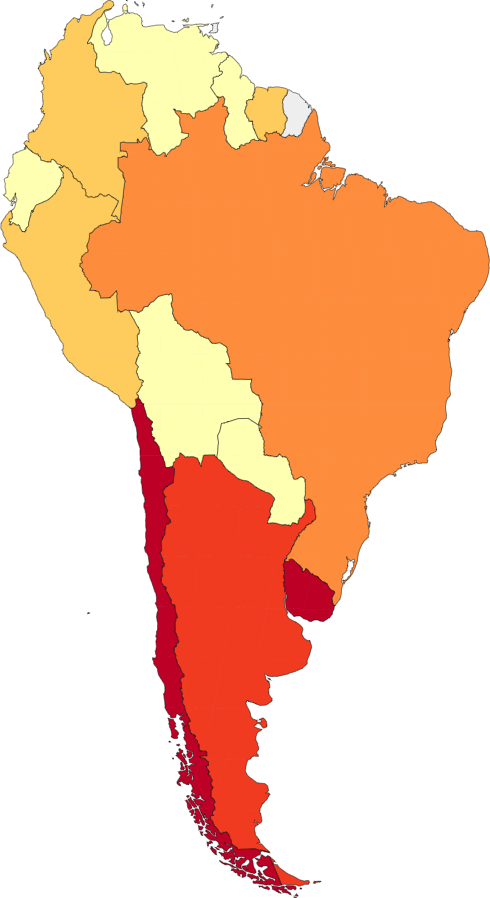}
    \caption{Equal Area}
    \label{fig:south-america-area}
  \end{subfigure}%
  ~
  \begin{subfigure}[t]{0.25\columnwidth}
    \raisebox{12mm}{
    \begin{tikzpicture}[scale=1/4]
        \legend[2](\large \texttt{\#ffffb2})(\large \texttt{\#fecc5c})(\large \texttt{\#fd8d3c})(\large \texttt{\#f03b20})(\large \texttt{\#bd0026});
    \end{tikzpicture}
    }
    \caption{Color progression}
    \label{fig:colorbrewer}
  \end{subfigure}
  \caption{Choropleth maps for GDP per capita in South America (July 2017)~\cite{central_intelligence_agency_country_2019}.\label{fig:south-america-gdp}}
\end{figure}

For example, consider Figure~\ref{fig:south-america-gdp} which contains two
choropleth maps of South America visualizing each countries GDP per capita as
of July 2017 \cite{central_intelligence_agency_country_2019}. The data for each
map is the same, but the classification method in
Figure~\ref{fig:south-america-length} is equal length while
Figure~\ref{fig:south-america-area} uses equal area.  For both maps we use five
colors on a progression from yellow to red. This progression is shown in
Figure~\ref{fig:colorbrewer}. The problem that equal area classification aims
to solve is that Figure~\ref{fig:south-america-length} is visually dominated by
the darkest red color. In contrast Figure~\ref{fig:south-america-area} is much
more evenly colored.


The equal area method assigns the colors so that the screen areas spanned by
each color are roughly the same.  This is achieved by first
choosing a projection, and then computing the areas of all regions and finding
the breaks (where to split between colors) that achieve this coloring.  In this
paper we show that this can be achieved by a number of algorithms of varying
complexity and visual effect.

In our discussion of the equal area method we point out the need for
considering all the possible breaks combinations which is a problem of
exponential complexity.  Therefore, we have to devise smart algorithms to find
the breaks in which we are interested.  In this paper, we focus on clustering the areas corresponding to the regions
into chunks with roughly equal total area, aiming to achieve an area-wise
evenly-colored choropleth map.  We derive a number of algorithms for its
computation paying close heed to their complexity.  We show that the equal area
method serves as a reasonable compromise between the drawbacks of the other
methods from a visual perspective which is also confirmed via the aid of a user
study.

It is interesting to observe that the equal area classification process is
related to the Cartogram method~\cite{Cartogram} in the sense that the
equal area method reflects the relative values of the data by varying the colors
of the regions.  On the other hand, the Cartogram method does this by varying
their displayed area while preserving their shape, hence not needing the aid of
color for differentiating between the classes.

The contributions of our paper are: Algorithms for computing equal area
classification including an optimal one that makes use of dynamic programming,
and a new classification method for coloring a choropleth map that enables
users to combine the benefits of the equal area and equal length methods using
a parameter termed a {\em W-score} that is analogous to the f-score from
information retrieval that combines precision and recall~\cite{Salt89}.

The rest of this paper is organized as follows. Section~\ref{sec-related-work}
reviews related work.  Section~\ref{sec-equal-area-algorithms} presents a
number of algorithms to compute the equal area classification method.
Section~\ref{sec-evaluation} contains the result of an evaluation of the
algorithms from a visual perspective as well as the results of a user study.
Section~\ref{sec-optimized-algorithms} uses the results of the user study to
devise an optimized algorithm that is a blend between the equal length and
equal area methods.  Section~\ref{sec-concluding-remarks} contains concluding
remarks and directions for future work.

The paper contains a number of maps all of which were rendered using the
D3 JavaScript library.  In addition, we used the library's capability to
compute a region's area in order to find the area in pixels that each
country occupies on the screen (as opposed to the actual area in km$^2$
that it occupies on the spherical surface of the Earth) because the areas
change when the projections change.  We use the Winkel-Tripel projection
in most of our world maps, as it is regarded as one of the best world
projections~\cite{Gold07}.

Our examples make use of a color progression to depict the data.  There are
many possible progression available which can be broadly classified as
qualitative, binary, sequential, or diverging~\cite{brewer_color_1994}. Using
the ColorBrewer~\cite{harrower03} system, we selected a sequential scheme that
blends hues to obtain a progression from yellow to red. The exact colors that we
used are shown in Figure~\ref{fig:colorbrewer}. We limited our progression to
five distinct colors as this is recommended as a reasonable maximum by
\citet{slocum_thematic_2009}.

\section{Related Work}
\label{sec-related-work}
Choropleth maps are a well studied and commonly used visualization technique
for the display of geographic information. Examples of their use include
\citet{jern_treemaps_2009} in which choropleth maps are used in combination
with tree maps~\cite{johnson_treemaps_1991} to visualize hierarchical
demographic statistics and \citet{lima_chorolibre_2019} who use choropleth maps
as a component in a system for the display of a similarly structured
hierarchical dataset. Prior analysis of choropleth classification techniques
such as \citet{brewer_evaluation_2002} has thoroughly investigated the effect
of various well known classification methods on response accuracy for questions
asked about data visualized on a map. Other work such as
\citet{zhang_quantifying_2017} explores the visual impact modifying the class
boundaries of a choropleth map with respect to visual clustering of entities.

In the remainder of this section, we briefly review techniques for choropleth
map data classification and provide a more detailed look at prior work on equal
area classification. We divide classification techniques into two broad
classes: those that are purely statistical and those that incorporate
geographic information into the classification.

\subsection{Statistical Classification}
The most common classification techniques do not use any geographic information
when constructing class boundaries. The simplest of these are equal intervals
and equal length (quantiles) classification. In equal intervals, each class
covers the same range of possible input values; similarly, equal length
classification constructs classes such that each class contains the same number
of elements~\cite{slocum_thematic_2009}.  Despite its simplicity, equal length
classification has been found to be superior to more complicated options on
some datasets~\cite{brewer_evaluation_2002}.

A more complicated and widely used technique is Jenks natural
breaks~\cite{jenks_error_1971,Jenk77}. This classification scheme attempts to
classify data values into different classes according to the breaks or gaps
that naturally exist in the data by minimizing the amount of variance between
elements in the same class~\cite{slocum_thematic_2009}.

A more recent example of a purely statistical classification scheme is the
head/tail breaks method~\cite{Jian13}. The technique is designed to deal with
distributions that have a heavy tail such as power laws.  In other words, it
recognizes the fact that there are far more objects of small magnitude than
objects of large magnitude.

\subsection{Geographic Classification}
There are methods other than equal area classification that incorporate
ancillary geographic information when deciding on class boundaries. This is
generally desirable since, being visualised on a map, geographic information is
relevant to readers.

An example of this are two measures of choropleth map accuracy proposed by
\citet{jenks_error_1971}: overview accuracy index (OAI) and boundary accuracy
index (BAI). OAI incorporates weighting for area into the accuracy measure used
by the standard Jenks natural breaks classification by making it more important
that variance is minimized for classes with a large total area. BAI, while
proposed by \citeauthor{jenks_error_1971}, was fully developed into a concrete
measure of accuracy by \citet{armstrong_using_2003}. BAI optimizes for large
differences between neighboring polygons that are not in the same class.

Following a similar motivation to that of BAI, \citet{chang_task_2018} present
a method that assigns classes such that geographically local extrema are easily
identified. Such extrema are polygons that have a higher or lower value
than all neighboring polygons. To make these polygons identifiable,
\citeauthor{chang_task_2018} attempt to find a classification where these
polygons are not in the same class as their neighbors.

Instead of aiming to make neighboring regions more distinguishable,
\citet{mcnabb_dynamic_2018} merge small neighboring polygons into larger
polygons that are homogeneously colored. While this approach makes it
impossible to distinguish between the constituent polygons, it reduces the
overall number of polygons on a finished map which can increase the
perceptibility of areal patterns. To make individual polygons visible when
necessary, the authors dynamically recompute this merged classification as
users manipulate an interactive map. By doing this, the user can zoom in on
areas of interest to view the individual polygons.

\citet{du_banded_2018} described a variant on the standard choropleth map for
the display of spatiotemporal data using a method they call the banded
choropleth map. In their method, each areal unit in the choropleth map is
further subdivided into a number of vertically oriented bands where each band
represents a discrete time interval. They present two methods for determining
the placement of these bands: one in which each band in a given areal unit has
the same width and another in which each band has the same area.  These are
analogous to the equal length and equal area classification methods in a
traditional choropleth map.

\subsection{Equal Area Classification}
Equal area classification incorporates geographic information, but we address
it separately since it is directly related to this paper. This classification
scheme has clearly been considered by researchers in the past, but there are
limited details available on past implementations. For instance,
\citet{murray_integrating_2000} make use of an implementation of equal area
classification provided by Esri ArcView~3.1, but public details are not
available on this implementation and the classification scheme was dropped from
later releases of the software.

\citet{robinson_elements_1984}\footnote{We deliberately reference the
\citeyear{robinson_elements_1984} edition of this book. Discussion of equal
area classification is absent in later editions.} address equal area
classification in their book and suggest an alternate name for the
classification technique: ``geographical quantiles''. This name emphasises
the similarity between equal area and equal length classification. To compute
an equal area classification, the authors suggest consulting a cumulative
frequency graph to find breaks after accumulating the desired percentage of
total area. Computing breaks by hand with this technique should yield results
that are very similar to the first greedy algorithm we propose in
Algorithm~\ref{alg:greedy1}.

The earliest treatment of equal area classification we were able to find is in
a series of papers by \citeauthor{lloyd_decision_1976} starting in
\citeyear{lloyd_decision_1976}
\cite{lloyd_decision_1976,lloyd_visual_1977,steinke_judging_1983}. In these
papers the authors evaluate the factors used by map readers when making comparisons
between different maps. They conclude that map readers first compare maps
according to their relative blackness before considering other map features. To
aid readers in comparing maps, they propose and evaluate using an equal area
classification to keep relative blackness constant between maps, forcing the
reader to compare other features. While equal area classification is discussed
at length in these papers, no specific technique for its computation is
mentioned.

\citet{armstrong_using_2003} obtain an equal area classification by minimising
a measure of areal inequality termed Gini coefficient for equal area (GEA)
using a genetic algorithm. While this approach obtains an equal area
classification, it is not able to make guarantees with respect run time or
optimality of the classification due to the use of genetic algorithms in
finding the solution. Rather than focusing specifically on equal area
classification this paper obtains classification schemes that are optimized
with respect to multiple error metrics using the aforementioned genetic
algorithms. This is similar in concept to the balanced classification we
propose in Algorithm~\ref{alg:dp-balanced}.  While the authors do not
demonstrate the specific equal area and equal length combination that we use,
their framework could, in principle, obtain such a classification.  A software
package that implements these algorithms is described in a later
paper~\cite{xiao_choroware_2006}.

Rather than dividing the map into some number of regions with approximately the
same area, it is possible to assign each class in the map a target proportion
of the total area and attempt to satisfy these proportions when creating class
boundaries. \citet{brewer_evaluation_2002} term this technique shared area
classification and incorporate it into their evaluation of choropleth map
classification techniques for use with epidemiological data. Prior to this
work, \citet{carr_hexagon_1992} applied the same classification scheme to
hexagon mosaic maps. In neither case is a concrete algorithm for computing of
the classification given.

\section{Equal Area Algorithms}
\label{sec-equal-area-algorithms}

In this section we give a definition of the equal area classification problem
and outline a number of algorithms to perform it of varying complexity.

\subsection{Problem Definition}
Given a sequence of $N$ numbers (the area of each region, sorted by the value
of the property associated with the regions), and a number $K$, the goal is
to partition the sequence into $K$ parts, so that the sum of the areas
comprising each part is roughly equal.


\subsection{Greedy Algorithms}
\label{sec-greedy-equal-area}

The simplest algorithm (Algorithm~\ref{alg:greedy1}) is to build the chunks
one-by-one while traversing the sequence of values in increasing order.  We
start inserting values into the chunk until the sum of the elements in the
chunk exceeds the target chunk sum, at which time we deem it full, and
start filling a new chunk.

\begin{algorithm}[!t]
\caption{Greedy algorithm for partitioning a sequence of $N$ numbers
  into $K$ chunks with roughly an equal sum. Complexity: $O(N)$.}
\label{alg:greedy1}
\begin{algorithmic}[1]
\Statex
\Function{GreedySplit}{$numbers$, $K$}
\Let{$average\_chunk$}{$\dfrac{\textrm{sum(}numbers\textrm{)}}{K}$}
\Let{$chunks$}{$<$empty list of chunks$>$}
\Let{$new\_chunk$}{$<$empty list of numbers$>$}
\For{$number$ in $numbers$}
	\State Append $number$ to $new\_chunk$.
	\If{sum($new\_chunk$) $\ge average\_chunk$}
		\State Append $new\_chunk$ to $chunks$.
		\Let{$new\_chunk$}{$<$empty list of numbers$>$}
	\EndIf
\EndFor
\State Append $new\_chunk$ to $chunks$.  \Comment{The last chunk.}
\State \Return{$chunks$}
\EndFunction
\end{algorithmic}
\end{algorithm}

Since the above greedy algorithm always overestimates the chunks (it keeps
inserting in the chunks until the sum of the elements in the chunk exceeds the
target chunk sum) the last chunk will be significantly smaller than the
others. To mitigate this issue, we can consider the total sum so far in order
to decide when to start a new chunk. This modified approach is presented in
Algorithm~\ref{alg:greedy2}.

\begin{algorithm}[!t]
\caption{Modified greedy algorithm for partitioning a sequence of $N$ numbers into $K$ chunks with roughly equal sum. Complexity: $O(N)$.}
\label{alg:greedy2}
\begin{algorithmic}[1]
\Statex
\Function{GreedySplit2}{$numbers$, $K$}
\Let{$average\_chunk$}{$\dfrac{\textrm{sum(}numbers\textrm{)}}{K}$}
\Let{$chunks$}{$<$empty list of chunks$>$}
\Let{$new\_chunk$}{$<$empty list of numbers$>$}
\Let{$num\_chunks$}{1}

\For{$number$ in $numbers$}
	\State Append $number$ to $new\_chunk$.
	\If{sum(numbers up to number) $\ge$ \\
		\hskip 4.1em $average\_chunk \times num\_chunks$}
		\State Append $new\_chunk$ to $chunks$.
		\Let{$new\_chunk$}{$<$empty list of numbers$>$}
		\State $num\_chunks$++
	\EndIf
\EndFor

\State Append $new\_chunk$ to $chunks$.  \Comment{The last chunk.}
\State \Return{$chunks$}
\EndFunction
\end{algorithmic}
\end{algorithm}

\subsection{Optimal Dynamic Programming Algorithm}
\label{sec-dynamic-programming-equal-area}

Before we present an optimal algorithm, let us define the criterion which we
are using to measure the performance of an algorithm that returns a set of $K$
chunks that partition the sequence of values:
$$ERROR\left( {breaks} \right) = \frac{1}{K} \sum\limits_{i = 1}^K {\left| {sum\left( {chunk_i} \right) - AVG} \right|}$$
$AVG = \dfrac{{sum\left( {values} \right)}}{K}$ is the sum of the
optimal chunk, and $sum(chunk_i)$ is the sum of the values in the
$i^{th}$ chunk of the partitioning. In other words, the formula computes the
average distance between the sum of a chunk in the given partitioning, and the
optimal chunk sum.  An optimal algorithm would find a set of breaks to
partition the list with the minimal {\em ERROR}.

\paragraph{Key Idea 1}

The first key idea in the algorithm, is that we can find an optimal position
for the last break with a linear pass on the values. We can start traversing
the list in decreasing order and sum up the values, until they add up to at
least {\em AVG} (let $x$ be the last value we include). Since the values are
arbitrary, it's unlikely that they will add up to {\em AVG} exactly, so the sum
will be slightly greater than {\em AVG} if we include $x$ in the final chunk
and slightly less than {\em AVG} if we do not include $x$ in the final chunk.

\textbf{Observation:}
There is an optimal partitioning in which the final break is around $x$ --
either before it or after it.
\textbf{Proof:}
Given an optimal partitioning of the list of values, if the final break is
around $x$, then we are done.  Otherwise, there are two options for the final
break: 

%
%
%
%

\begin{enumerate}
\tightlist
\item The final break is before candidate 1. In this case, we argue that
    moving it to where candidate 1 is can only reduce the error. Notice that
    moving the final break to the right (assuming that the list of values is
    sorted in increasing order from left to right) modifies only the final two
    chunks, call them $A$ and $B$. The sum of chunk $A$ will increase by some
    $\delta$ (the sum of the values between the   current break and candidate
    1), and the sum of chunk $B$ will  decrease by the same $\delta$. Since the
    sum of chunk A was greater than $AVG$ and is still greater, but now is
    $\delta$ closer to $AVG$, the error stemming from it will decrease by
    exactly $\frac{\delta}{K}$. Since the sum of chunk B changed by
    $\delta$, the error stemming from it can increase by at most $\frac{\delta}{K}$.
    Therefore, the error in the new partitioning can only
    decrease. Hence, the new partitioning is also optimal.

\item The final break is after candidate 2. In this case, we argue that
    moving it to where candidate 2 is can only reduce the error. Similarly to
    the other case, the error from the final chunk will decrease by exactly
    $\frac{\delta}{K}$. The only caveat here is that in the given
    optimal partitioning, there might be multiple breaks after candidate 2. In
    this case, we will move them one by one to where candidate 2 is, starting
    from the leftmost one (i.e., having the smallest value). Every time we
    shift a ${\it break}_i$ to the left, the error from the chunk on its right
    will decrease by some $\frac{\delta_i}{K}$, and the error from the
    chunk on its left can increase by at most $\frac{\delta_i}{K}$.
    Therefore, the total error cannot grow. Hence, the new partitioning is also
    optimal.
\end{enumerate}
$\qed$

The pseudocode for a procedure to find the two candidates described above is
given below.
\begin{algorithmic}[1]
\Statex
\Function{FindLastBreakCandidates}{$values$}
\Let{$s$}{$0$}  \Comment Sum of the values so far.
\For{$i = N - 1$ until $0$}
	\Let{$s$}{$s + values[i]$}
	\If{$s > AVG$} \State \Return{($i$, $i+1$)} \EndIf
\EndFor
\EndFunction
\end{algorithmic}

\paragraph{Key Idea 2}

Let $BestError(m, b)$ be the minimum error we can get for placing $b$ breaks to
partition the first $m$ elements of the list (which has a total of $N$ values),
and $Breaks(m, b)$ be the set of corresponding breaks. We are interested in
$Breaks(N, K - 1)$.

Building on top of Key Idea 1, we give the following recursive definition of
$BestError$. First, find the two candidate locations for the final pivot,
call them $l_1, l_2$. Next compare $BestError(l_1, p - 1) +
|\Sigma(\textrm{values between } l_1 \textrm{ and } m) - AVG|$ and
$BestError(l_2, p - 1) + |\Sigma(\textrm{values between } l_2 \textrm{ and } m)
- AVG|$. Select the break with the smaller error.

In other words, for each candidate for the location of the final
break, we recursively find the other breaks, and then compare the error from
the two options to choose the better candidate.  Calculating this recursively
will have a high computational cost, as we might repeat many computations.
Therefore, we will compute it using dynamic programming.  In particular, we
allocate a two-dimensional array to store the values of $Breaks(m, b)$ for all
$m \in [0, N], b \in [0, K - 1]$, and compute them column-by-column so that
whenever we need a value for $Breaks(loc_{1 \textrm{ or }2}, b - 1)$ we can
look it up in the table in constant time.

\paragraph{Key Idea 3}

As described in Key Idea 1, finding the candidates for the final break takes
linear time, as we need to traverse the list of values in decreasing order.
However, as we are using dynamic programming to compute all the values for
$Breaks(m, b)$ (for all values of $m$ and $b$), we can save on some
computations.  Notice that if the candidates for the rightmost break in
$Breaks(m, b)$ were around position $x$, then the candidates for the rightmost
break in $Breaks(m - 1, b)$ have to be on the left of $x$ (they can't be on the
right), because the list is shrinking from the right, so the rightmost chunk
needs to grow from the left to remain close to the average.  Thus the break
needs to move left.  Therefore, we can compute the $Breaks(m, b)$ in a loop
where we move $m$ and the candidates backward simultaneously until they hit
index 0.  This way we compute $N$ cells in the $Breaks(m, p)$ array in $O(N)$
time.

\paragraph{Key Idea 4}

In the above key ideas we used the \textit{sum} function, which when
implemented naively, requires linear time. By preprocessing the sequence of
values and building a cumulative sum array (e.g. given $[a, b, c]$ construct
$[0, a, a+b, a+b+c]$), we can build a function that returns the sum of any
sub-sequence between indices $i$ and $j$ in constant time. This procedure is
presented in Algorithm~\ref{alg:psum}.

\begin{algorithm}[!t]
\caption{Prefix sum array calculation. \textsc{Preprocess}($values$) is invoked
    once on the original array, to compute the prefix sum array in $O(N)$ time.
    After that, \textsc{PSum}($i$, $j$) can be invoked to compute the sum of
    the values between any two indices $i$ (inclusive) and $j$ (exclusive) in
the original array in $O(1)$ time.}
\label{alg:psum}
\begin{algorithmic}[1]
\Statex
\Function{Preprocess}{values}
\Let{$sums\_array$}{$[0]$}
\Let{s}{$0$}
\For{$x$ in $values$}
	\Let{$s$}{$s + x$}
	\State Append $s$ to $sums\_array$.
\EndFor
\State \Return $sums\_array$
\EndFunction
\end{algorithmic}
\begin{algorithmic}[1]
\Statex
\Function{PSum}{$i$, $j$}
\State \Return $sums\_array[j] - sums\_array[i]$
\EndFunction
\end{algorithmic}
\end{algorithm}

It is important to note that from now on, when we invoke \Call{PSum}{i,
j} in the pseudocodes, then we assume that the preprocessing above has
been performed, and we call \Call{PSum}{i, j} to calculate the sum of the
values between \textit{indices} i (inclusive) and j (exclusive) in
constant time.  Indices are 0-based.  Using the above ideas,
Algorithm~\ref{alg:equal-dp} gives the pseudocode for the optimal
algorithm to find breaks that partition a sequence of $N$ values into $K$
chunks with a roughly equal sum.

\begin{algorithm}[!t]
\caption{Dynamic Programming optimal algorithm for finding the breaks for partitioning a sequence of
    $N$ values into $K$ chunks with a roughly-equal sum. This algorithm returns
    the breaks. Complexity: $O(N K)$.}
\label{alg:equal-dp}
\begin{algorithmic}[1]
\Statex
\Function{DPOptimalEqualArea}{values}
\Let{$AVG$}{sum($values$) / $K$}  \Comment Average chunk sum.
		
\Let{$best\_error$}{
$<$\parbox[t]{\dimexpr 17em}{
2D array with $N+1$ rows (from 0 to $N$) for the end index $m$, \\
and $K$ columns (from 0 to $K-1$) for the number of breaks $b$.$>$}
}
\Let{$best\_breaks$}{
$<$\parbox[t]{\dimexpr 16em}{2D array like above, for the breaks$>$}
}
\For{$m$ in $0..N$} \Comment Fill the first column (column 0).
	\Let{$best\_error$[$m$][$0$]}{$\lvert$\Call{PSum}{$0$, $m$}$ - AVG$$\rvert$}
	\Let{$best\_breaks$[$m$][$0$]}{[]}
\EndFor

\For{$b$ in $1..K-1$}   \Comment Loop over breaks.
	\Let{$m$}{$N$}
	\Let{$break$}{$N$}  \Comment The position of the final break.
	\While{$m \ge 0$}
		\If{$break > m$}
			\Let{$break$}{$m$}
		\EndIf
		
		\State \Comment \parbox[t]{\dimexpr 20.35em}{Go back until reaching the candidate positions for the last break:}
		\While{\Call{PSum}{$break$, $m$}$ < AVG \land  break > 0$}
			\Let{$break$}{$break - 1$}
		\EndWhile
		
		\State \Comment \parbox[t]{\dimexpr 20.35em}{Choose between the two candidates for the last break:}
		\If{$best\_error$[$break + 1$][$b-1$]
			    \NoNumber{+ $\lvert$\Call{PSum}{$break + 1$, $m$}$- AVG$$\rvert <$}
			    \NoNumber{$best\_error$[$break$][$b-1$]}
			    \NoNumber{+ $\lvert$\Call{PSum}{$break$, $m$}$- AVG$$\rvert$}}
			\Let{$break$}{$break + 1$}
		\EndIf
		\State \Comment \parbox[t]{\dimexpr 20.35em}{After choosing the better break, add it to the arrays.}
		\Let{$best\_error$[$m$][$b$]}{best\_error[$break$][$b-1$]
					\NoNumber{\tab\tab\tab\tab + $\lvert$\Call{PSum}{break, m}$- AVG$$\rvert$}}
		\Let{$best\_breaks$[$m$][$b$]}{$best\_breaks$[$break$][$b-1$] + [$break$]}
		\Let{$m$}{$m - 1$}
	\EndWhile
\EndFor
\State \Return $best\_breaks$[$N$][$K - 1$]
\EndFunction
\end{algorithmic}
\end{algorithm}

\subsubsection{Computational Complexity}

Algorithm~\ref{alg:equal-dp} fills a table with $O(N)$ rows and $O(K)$
columns, and takes constant time to fill each cell. Therefore, the total
space and time complexity is $O(N K)$, which is almost linear, as $K$ is
usually low (3--10 colors on the map).  Note that filling each column
(with $N$ cells) takes $O(N)$ time, because as explained in Key Idea 3,
the variables $m$ and $break$ which control the nested loops in lines 12
and 17 of the pseudocode, start from $N$ and go backwards together until
reaching 0.  In other words, these two nested while loops run in linear
time.

As written, the algorithm requires $O(N K^2)$ space because
$best\_breaks[m][b]$ stores $b$ breaks. However, this can be easily optimized
so that $best\_breaks[m][b]$ would store only the last break and a pointer to
the list of the other breaks (stored in a different cell). The full list of $b$
optimal breaks would then be retrieved at the end of the computation.  This way
the space complexity will be $O(N K)$.

\subsubsection{Caveat - Using a Different Optimization Criteria}

At the start of this section, we defined the optimality criterion to be the
difference in absolute value from the average chunk sum. An alternative
optimality criterion could be formed by the sum of squares of the differences.
In this case, key idea 1 will no longer hold, as shifting the breaks has a
different effect on the squared errors. Therefore, we cannot only consider two
candidate locations for the final break. Instead, we will need to consider all
the possible locations. This will require linear time for computing every cell
in the two-dimensional array, leading to a total complexity of $O(N^2 K)$.

\section{Evaluation}
\label{sec-evaluation}

Section~\ref{sec-equal-area-algorithms} described a number of variations of the
equal area algorithm.  They differed in part on the basis of their
computational complexity and on the extent to which the areas of the resulting
chunks differed from being equal which is characterized by the term {\em
error}.  In this section we use a dataset to evaluate the different equal area
algorithms as well as their alternatives which are the naive equal length and
the Jenks natural breaks methods.  In the case of the equal area algorithms,
the evaluation is relative and is in terms of both low quantitative error and
execution time perspectives.  In the case of the alternative methods, the
evaluation is from both a visual perspective and a user study, where the
equal area algorithm with the lowest error is used.

The rest of this section is organized as follows.
Section~\ref{sec-equal-area-eror-comparison} evaluates the maps that result from the
application of the various equal area algorithms to a dataset.  Having chosen
the equal area algorithm that achieves the lowest quantitative error,
Section~\ref{sec-visual-comparison} uses a visual perspective to compare its use on
the same dataset with the alternative methods using both the Winkel-Tripel and
Mercator projections.  Section~\ref{sec-user-studies} repeats the comparison with
the aid of a user study. Section~\ref{sec-density-maps} performs the comparison on
an alternative related dataset which provides the motivation for a generalized
algorithm (Section~\ref{sec-optimized-algorithms}) that combines the benefits of
both the equal area and equal length methods.

Before proceeding further, we first describe the dataset used in our
evaluation. It consists of population by country mapped using the Winkel-Tripel
projection. Choice of projection is particularly important because we compute
equal area breaks using projected area in pixels on the screen rather than real
land area. Using a projection that makes no attempt to minimize area distortion
could yield considerably different classifications that do not accurately
represent geographic distribution.

\subsection{Equal Area Algorithm Error Comparison}
\label{sec-equal-area-eror-comparison}
Table~\ref{tab:average-error} contains a summary of the average errors
resulting from implementing a number of variations of the equal area algorithm
given in Section~\ref{sec-equal-area-algorithms} for our world population
dataset with the Winkel-Tripel projection where the goal was to reduce the
average error with the baseline so that the areas assigned to each chunk or
color were equal.  Notice that for our dataset, using the first greedy
algorithm to assign the colors almost halves the error, and using the second
greedy algorithm shrinks it even further.  Moreover, using the optimal dynamic
programming algorithm produces a significantly better result than all the
others. Since the execution time complexity of the optimal dynamic programming algorithm is
good (almost linear), we recommend using it instead of the simpler algorithms
and this is the variant of the algorithm that we use in the comparisons
described in the remaining sections.

\begin{table}
\caption{Average difference between the optimal sum and the sum of each chunk.}
\label{tab:average-error}
\scriptsize
\centering
\begin{tabular}{ll}
\toprule
\textbf{Algorithm} & \textbf{Average Error} \\
\midrule
Equal length (Quantiles) & 34,928 \\
Greedy algorithm 1 & 15,192 \\
Greedy algorithm 2 & 5,572 \\
Optimal dynamic programming & 3,244 \\
\bottomrule
\end{tabular}
\end{table}

\subsection{Visual Comparison}
\label{sec-visual-comparison}

\begin{figure*}[t!]
\centering
   \begin{subfigure}[t]{0.7\columnwidth}
	 \includegraphics[width=\columnwidth]{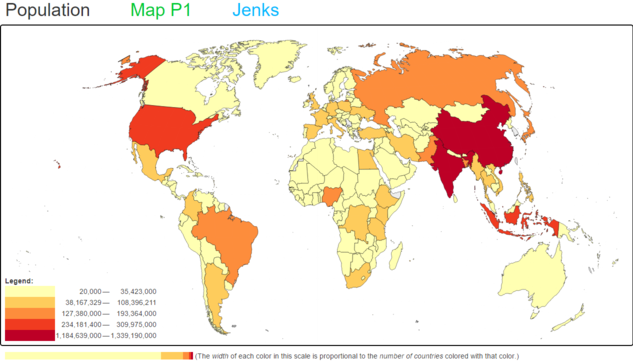}
     \caption{Natural Breaks --- Winkel-Tripel}
     \label{fig:p1}
   \end{subfigure}
   ~
   \begin{subfigure}[t]{0.7\columnwidth}
     \includegraphics[width=\columnwidth]{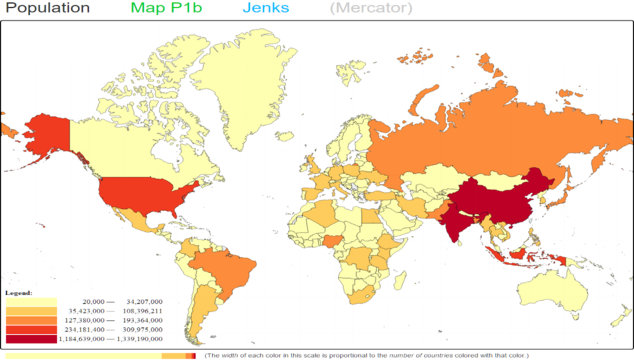}
     \caption{Natural Breaks --- Mercator}
     \label{fig:mapp1}
   \end{subfigure}

   \begin{subfigure}[t]{0.7\columnwidth}
	 \includegraphics[width=\columnwidth]{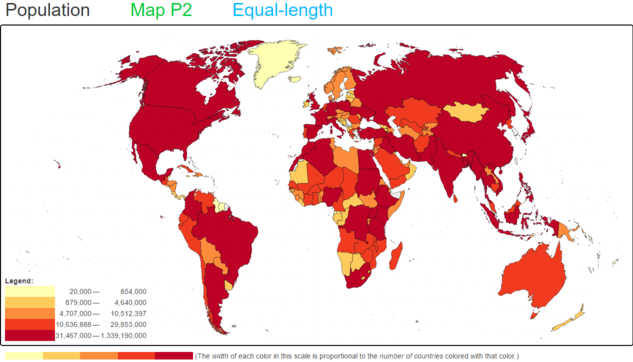}
     \caption{Equal length (Quantiles) --- Winkel-Tripel}
     \label{fig:p2}
   \end{subfigure}
   ~
   \begin{subfigure}[t]{0.7\columnwidth}
     \includegraphics[width=\columnwidth]{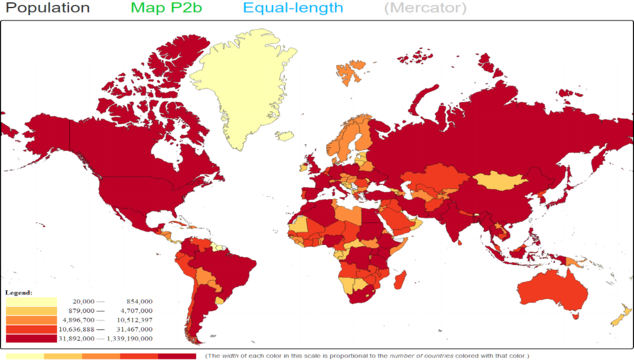}
     \caption{Equal length (Quantiles) --- Mercator}
     \label{fig:mapp2}
   \end{subfigure}

   \begin{subfigure}[t]{0.7\columnwidth}
	 \includegraphics[width=\columnwidth]{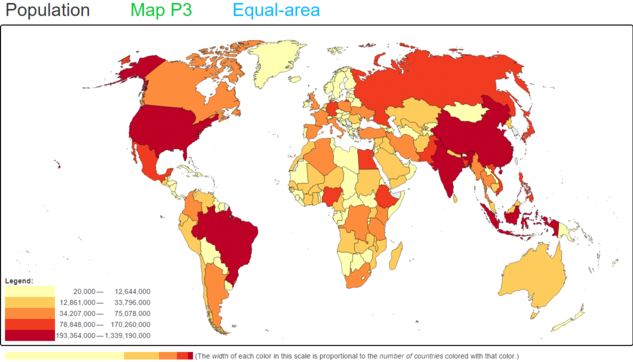}
     \caption{Equal area --- Winkel-Tripel}
     \label{fig:p3}
   \end{subfigure}
   ~
   \begin{subfigure}[t]{0.7\columnwidth}
     \includegraphics[width=\columnwidth]{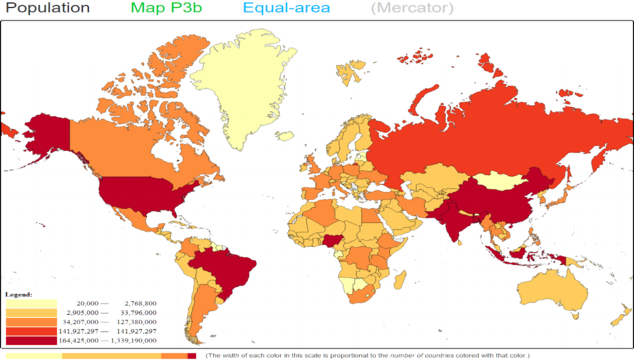}
     \caption{Equal area --- Mercator}
     \label{fig:mapp3}
   \end{subfigure}
\caption{Equal area compared to two common choropleth classification
         methods on Mercator and Winkel-Tripel projections.}
\label{fig:examples}
\end{figure*}

In addition to the error and computational complexity evaluation in
Section~\ref{sec-equal-area-eror-comparison} we also created maps
resulting from the use of the Jenks natural breaks, equal length, and the
optimal dynamic programming equal area algorithms using the Winkel-Tripel
projection, and they are given in Figures~\ref{fig:p1}, \ref{fig:p2},
and~\ref{fig:p3}, respectively.  Each map contains a legend that indicates the
range of data values for each color.  In addition, each map contains a scale at
its bottom that indicates for each color the proportion or number of the
countries that are depicted in it.


From the images we see that the lighter colors (corresponding to smaller areas
which generally also have lower populations) dominated when using the Jenks
natural breaks algorithm, while the darker colors (corresponding to larger
areas which generally also have higher populations) dominated when using the
equal length algorithm.  On the other hand, the color distribution area-wise is
clearly superior when using the equal area algorithm to the other two methods.
This is primarily because the equal area algorithm reduces the number of highly
populated and highly sized countries for the darkest color.  It also increases
the number of smaller populated and smaller sized countries for the lightest
color.  The natural breaks are hard to predict other than to note that they are
far less likely to occur for the countries with a small population which is why
the area of the countries with the lighter color is much greater when using the
Jenks natural breaks algorithm vis-a-vis the equal length algorithm.  On the
other hand, the difference in the area of the countries with the lighter color
in the Jenks natural breaks algorithm from those in the equal area method is
much smaller than the difference from those in the equal length method.

The above comparison used the Winkel-Tripel projection because it is an
equal area projection and hence thought to be more relevant to the goal of our
study which was to demonstrate the utility of the equal area method.  It turns
out that the nature of the projection did not have a material effect on our
evaluation except for the equal area projection.  In order to see the relative
independence of our comparison from the chosen projection,
Figures~\ref{fig:mapp1}, \ref{fig:mapp2}, and~\ref{fig:mapp3}, show Choropleth
maps when using the Mercator projection for the Jenks natural breaks,
equal length, and equal area methods, respectively.  We observe that the
Mercator projection has almost identical behavior in terms of the color
assignment for the Jenks natural breaks and equal length methods as does the
Winkel-Tripel projection although the rationale for using the projections is
different.  In particular, the rationale for using the Winkel-Tripel projection
is that it attempts to minimize area distortion. The Mercator projection
greatly distorted areas for countries near the poles (overestimation) and the
Equator (underestimation).  This distortion is the reason for the difference
between the projections for the equal area algorithm where we see that the
distortion (exaggeration) in the area of Greenland causes a reduction in the
number of the lightest colored countries in the Mercator projection vis-a-vis
the number in the Winkel-Tripel projection.  This can be observed by looking at
the African continent where the number of the countries with the lightest color
is lower in the Mercator projection than it is in the Winkel-Tripel projection.

\subsection{User study}
\label{sec-user-studies}

To evaluate the utility of the equal area classification algorithm for the
coloring of choropleth maps, we conducted a survey of 30 arbitrary people
through Amazon Mechanical Turk.  Participants were shown four maps of the
continental United States containing the same data classified using four
different techniques (equal area, equal length, equal interval, and natural
breaks), but participants were not informed that underlying data for each map
was the same.  They were then asked questions about the underlying data for the
maps to judge how effective each map was at conveying patterns in the data to
readers of the map. Finally, participants were asked to make a subjective
comparison between the four maps by indicating which map they felt was most and
least visually appealing. Similar task based approaches to evaluating
choropleth maps have been used in prior
studies~\cite{brewer_evaluation_2002,du_banded_2018,schiewe_empirical_2019}.

Before asking participants about the contents of the maps, we gauged their
background knowledge in cartography and in the geography of the United States by
asking them to rank their familiarity with these areas as either ``not familiar'',
``slightly familiar'', or ``highly familiar''. For both questions, few
participants ranked themselves as not familiar. Responses to these questions
are tabulated in Table~\ref{tab:background}.

\begin{table}
\caption{User study participant background familiarity with cartography and United States geography.}
\label{tab:background}
\centering
\begin{tabular}{lrrrrrr}
\toprule
& United States Geography & Cartography \\
\midrule
Not Familiar & 2 & 5\\
Slightly Familiar & 16 & 22\\
Highly Familiar & 12 & 3\\
\bottomrule
\end{tabular}
\end{table}

The data we chose to use in creating the maps was the number of confirmed
COVID-19 cases per $100,000$ residents in each state as reported by the United
States Centers for Disease Control and Prevention on April 22, 2020.
Participants were not informed that this was the datasets because we did not
want prior knowledge of the distribution of COVID-19 cases to affect responses.
Even though higher resolution data is available (e.g.~county or zip code),
state level data was chosen because states have considerably greater variation
in area than any smaller division. The resulting classification from applying
the equal area algorithm approaches that obtained from equal length as the
variation in area between objects in the dataset approaches zero. An evaluation
of equal area classification under such circumstances would not be interesting,
as the results would be the same as any prior evaluation of equal length
classification (e.g.~\citet{brewer_evaluation_2002}).

Choropleth maps are a lossy abstraction of the data used in their creation, so
we did not expect participants to be able to make precise statements about
data values in individual states. Instead, we asked participants to make
comparative generalizations about the relative data values in different regions
of the map. In particular, we had participants compare the western (W), midwestern (MW),
northeastern (NE), and southern (S) regions of the United States as defined by the
Census Bureau. We did not expect all participants to be familiar enough with
United States geography to fully understand these regions, so we included in
our survey a map indicating exactly what states are included in each region.

The three questions we asked for each map are the following: (1) what region of
the country has the highest average value, (2) what region of the country has
the lowest average data value, and (3) compare the average data value in the midwest
to that in the south. All questions were posed multiple chose questions. The
possible answers for the first two were the regions is enumerated above.
Options for the third questions were that the average value in the midwest is
greater than (GT), less than (LT), or equal to (EQ) the average value in the
south. The correct answers to the questions are northeast, west, and that
the average value in the is less than that in the south respectively.

\begin{table*}
\caption{Responses to questions posed in our user study for four choropleth map
classification techniques. Bold entries indicate correct responses.} \label{tab:responses}
\centering
\begin{tabular}{l|rrrr|rrrr|rrrr|rrrr}
\toprule
& \multicolumn{4}{c|}{Equal Area} & \multicolumn{4}{c|}{Equal Interval} & \multicolumn{4}{c|}{Natural Breaks} & \multicolumn{4}{c}{Equal Length}\\
& MW & NE & S & W & MW & NE & S & W & MW & NE & S & W & MW & NE & S & W\\
\midrule
Question 1 & 5 & {\bf 17} & 5 & 3
           & 4 & {\bf 17} & 6 & 3
           & 6 & {\bf 19} & 3 & 2
           & 7 & {\bf 18} & 3 & 2\\
Question 2 & 9 & 5 & 6 & {\bf 9}
           & 8 & 8 & 3 & {\bf 11}
           & 7 & 8 & 3 & {\bf 12}
           & 10 & 7 & 5 & {\bf 8}\\
& EQ & GT & LT &  & EQ & GT & LT &  & EQ & GT & LT &  & EQ & GT & LT & \\
Question 3 & 14 & 6 & \bf{10} &
           & 14 & 8 & \bf{8} &
           & 19 & 8 & \bf{3} &
           & 12 & 10 & \bf{8}\\
Percent Correct & \multicolumn{4}{c|}{40}
                & \multicolumn{4}{c|}{40}
                & \multicolumn{4}{c|}{37.78}
                & \multicolumn{4}{c}{37.78}\\
Most Appealing  & \multicolumn{4}{c|}{15}
                & \multicolumn{4}{c|}{4}
                & \multicolumn{4}{c|}{3}
                & \multicolumn{4}{c}{8}\\
Least Appealing & \multicolumn{4}{c|}{5}
                & \multicolumn{4}{c|}{12}
                & \multicolumn{4}{c|}{7}
                & \multicolumn{4}{c}{6}\\
\bottomrule
\end{tabular}
\end{table*}

In Table~\ref{tab:responses} we tabulate participant responses to our
questions. We found that overall response accuracy was very similar across
all the classification techniques. Equal area and equal interval tied for the
highest average percent correct while natural breaks and equal length followed
very closely behind. This shows that the choice of classification method is not
overly important for accurately conveying patterns in data, at least for this
specific dataset.

If this is the case, then it is reasonable to pick a classification based on
more subjective criteria, such as how visually appealing it is. In the last two
questions of our survey, we had participants indicate which maps they found
most and least appealing. Out of 30 responses, 15 listed equal area as the most
appealing map while only five listed it as their least favorite.  Thus we have
clear evidence for a preference for the equal-area classsification.

\begin{figure*}[t!]
\centering
  \begin{subfigure}[t]{0.7\columnwidth}
     \includegraphics[width=\columnwidth]{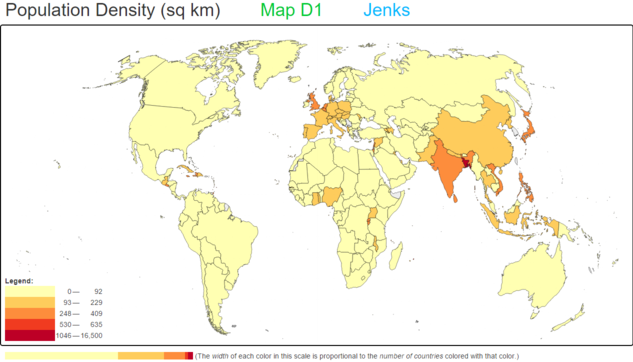}
     \caption{Natural Breaks}
     \label{fig:d1}
  \end{subfigure}
  ~
  \begin{subfigure}[t]{0.7\columnwidth}
    \includegraphics[width=\columnwidth]{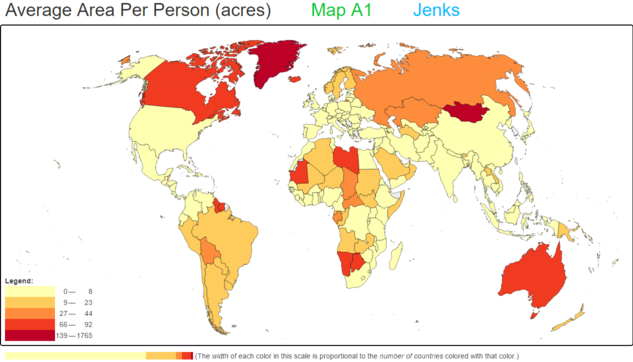}
    \caption{Natural Breaks}
    \label{fig:a1}
  \end{subfigure}

  \begin{subfigure}[t]{0.7\columnwidth}
    \includegraphics[width=\columnwidth]{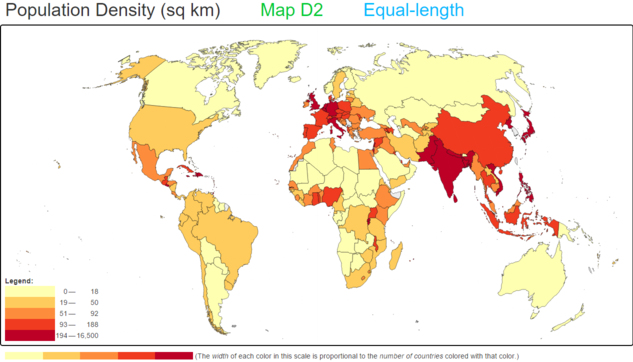}
    \caption{Equal length (Quantiles)}
    \label{fig:d2}
  \end{subfigure}
  ~
  \begin{subfigure}[t]{0.7\columnwidth}
    \includegraphics[width=\columnwidth]{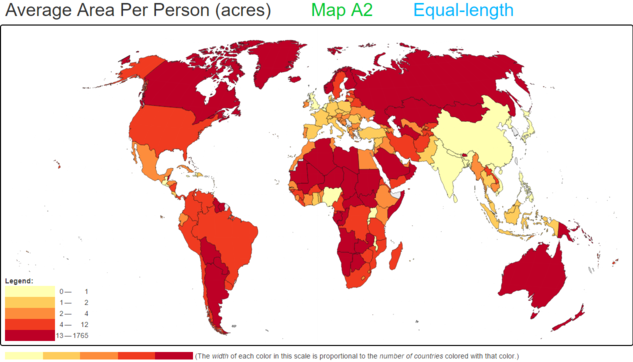}
    \caption{Equal length (Quantiles)}
    \label{fig:a2}
  \end{subfigure}

  \begin{subfigure}[t]{0.7\columnwidth}
    \includegraphics[width=\columnwidth]{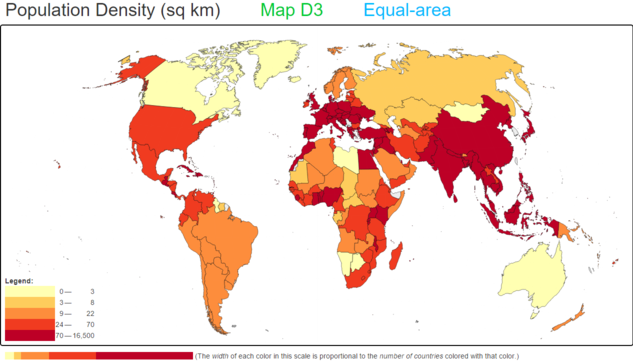}
    \caption{Equal area}
    \label{fig:d3}
  \end{subfigure}
  ~
  \begin{subfigure}[t]{0.7\columnwidth}
    \includegraphics[width=\columnwidth]{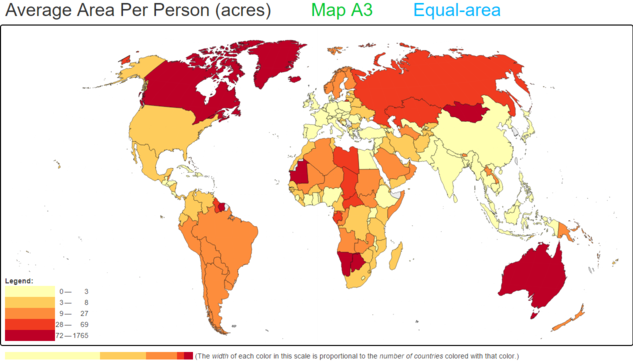}
    \caption{Equal area}
    \label{fig:a3}
  \end{subfigure}
\caption{Population Density and Area Per Person Maps}
\end{figure*}

\subsection{Density Maps}
\label{sec-density-maps}
Our discussion of the advantages of the equal area method has been in the
context of Choropleth maps.  It is interesting to note that there is a school
of thought that claims that a common error in the use of Choropleth maps is
their use to visualize raw data such as population where colors serve to
differentiate between magnitudes rather than to visualize normalized  values
such as densities where the colors are more intuitively used to convey
intensities.  This is based on the natural tendency to associate darker colors
with larger intensities and lighter colors with lower intensities.  The use of
a normalized measure such as density reflects the distribution of a value over
an area.  Thus in the case of raw data values, it is suggested that
proportional symbol maps are used instead of
choropleth~\cite{slocum_thematic_2009}.

This is fine when the areas are all of approximately the same size, but it is
not appropriate for areas that differ greatly in size as is the case with a map
of countries which are irregularly shaped. In this case point symbols are
likely to overlap and thus greatly mask the different areas.  Approaches such
as necklace maps have been suggested to resolve this particular issue while
continuing to use proportional symbols~\cite{speckmann_necklace_2010}.  It is
also worth noting that the equal area method is a form of normalization where
the normalization is applied to the embedding space (i.e., the world map) to
yield equal-sized areas, while density measures normalize the measured quantity
over an area.

Despite the above we still feel that the use of Choropleth maps for the
visualization of raw data such as population is important even if conventional
cartographic wisdom frowns on their use and calls instead for the use of
densities.  Our aim in this paper is to introduce the equal area method to
novices who form the majority of users rather than just experienced
cartographers, as these novices are more likely to be the producers of the maps
that are so common on the web today.  Therefore, it is important for us to make
sure that the coloring algorithm used in Choropleth maps produce good results
even if the data is not completely appropriate for being visualized with them.
This is what motivated our development of the equal area algorithm.
Figures \ref{fig:d1}, \ref{fig:d2}, and \ref{fig:d3} show Choropleth maps for
the population density using the Winkel-Tripel projection for the Jenks natural
breaks, equal length, and equal area methods, respectively.

Although not shown explicitly here, we note that most of the countries have
similar population densities.  This means that it is hard to find natural
breaks for the smaller population densities, while it is easier to find the
natural breaks for the larger population densities.  This is why most of the
countries are colored lightly (i.e., yellow) when using the Jenks natural
breaks algorithm and only India, China, Indonesia, and a few European countries
colored darkly.  The natural breaks are simply too few.  We also observe that
the larger countries tend to have smaller population densities which is why in
the equal length algorithm a large area is colored yellow although much less often
than with the Jenks natural breaks algorithm but still substantially less often than
the area colored yellow in the equal area algorithm where the total areas of
the countries associated with each color is approximately the same.

One way to overcome the drawback associated with the use of the Jenks natural
breaks and equal length algorithms for population density is to note that a
more appropriate measurement alternative to population density is the inverse
measurement of average area per person.  Figures \ref{fig:a1}, \ref{fig:a2},
and \ref{fig:a3} show Choropleth maps for the average area per person using the
Winkel-Tripel projection for the Jenks natural breaks, equal length, and
equal area methods, respectively.  In this case, we find that fewer of the
countries have similar average area per person but this number is still
substantial and thus its still hard to find natural breaks (e.g., some of the
countries like the US are still colored yellow as also in the case of the
density maps).  Nevertheless there is greater variation in the colors of the
countries.  Taking the inverse measurement of average area per person makes no
difference in the map produced by the equal length algorithm other than making
the formerly lightly colored countries dark and making the formerly darkly
colored countries light.  In the example map, we find that since the larger
countries tend to have higher values of average area per person, a large part
of the map will be colored darkly.  The fact that the total areas of the
countries associated with each color is approximately the same for the
equal area algorithm means that when using it there is no difference in the color
distribution for the population density measurement or its inverse
measurement other than the change in the color associated with each country
from light to dark and from dark to light.

\section{Optimized Algorithms}
\label{sec-optimized-algorithms}
Recall that each map contains one legend for the boundaries of data values for
each color, and another for the proportion of countries associated with each
color.  This information indicates the extent to which each map deviates from
the equal length map in the sense of whether the number of countries displayed
with that color deviates from being the same.  Looking at this information we
notice that the number of countries colored red (the darkest color) is usually
small.  However, this is not the case for the equal area population density map
where over half of the countries are colored red thereby covering almost all of
Europe and Asia leading to a perception that the map is not balanced even
though the colors are equally distributed in terms of area.  This prompted us
to try to improve our maps further, by attempting to also take into account the
number of countries that are assigned to each color.  In other words, we are
attempting to strike a better balance between the equal area classification and
the equal length classification.

\subsection{Optimized Greedy Algorithm}
\label{sec-optimized-greedy-algorithm}
\begin{algorithm}[!t]
\caption{Greedy algorithm for partitioning a sequence of $N$ numbers into $K$
chunks with roughly equal sum, while simultaneously trying to balance the
lengths of the chunks as well. The algorithm returns the breaks. Complexity: $O(N)$.}
\label{alg:balanced-greedy}
\begin{algorithmic}[1]
\Statex
\State U = 2  \Comment \parbox[t]{\dimexpr 19em}{Upper bound on the chunk length/sum ratio. Chunk length and sum should not exceed twice those of the average chunk.}
\State L = $\frac{1}{2}$  \Comment \parbox[t]{\dimexpr 19em}{Lower bound on the chunk length ratio. Chunk length should be at least half the length of the average chunk.}

\Function{OptimizedGreedy}{$numbers$,$start\_index$,$num\_chunks$}.
	\If{$num\_chunks$ == 1}
		\State \Return []  \Comment One color - no breaks.
	\EndIf
	\Let{$AVG$}{$\dfrac{\textrm{sum}(numbers)}{num\_chunks}$}	
	\Let{$AVG\_LEN$}{$\dfrac{len(numbers) - start\_index}{num\_chunks}$}
	\Let{$len$}{0}  \Comment The length of the new chunk.
	\Let{$s$}{0}  \Comment The sum of the new chunk.
    \For{$i$ in $start\_index$ .. len($numbers$)}
    	\Let{$len$}{$len + 1$}
    	\Let{$s$}{$s + numbers[i]$}
		\If{$(s \ge AVG$ $\land$ $len \ge L \times AVG\_LEN)$ $\lor$
           \NoNumber{$(len \ge U \times AVG\_LEN)$ $\lor$ $(s \ge U \times AVG)$}}
		   \State \Return \parbox[t]{\dimexpr 18em}{[i + 1] + \Call{OptimizedGreedy}{$numbers$, $i + 1$, $num\_chunks - 1$}}
		\EndIf
	\EndFor
\EndFunction
\end{algorithmic}
\end{algorithm}

In our construction of an equal area algorithm we proposed two greedy
algorithms (Section~\ref{sec-greedy-equal-area}).  In the first of these algorithms,
we added elements to a chunk until its sum exceeded the average area per chunk
denoted by {\em AVG}.  At this point, we make a few modifications so that
we can achieve a balance in the lengths of the chunks in addition to balancing
the sums of the areas of their elements.  In particular, we keep inserting
elements into a chunk until one of the following conditions is satisfied.
\begin{enumerate}
\tightlist
\item The chunk is full in terms of area (as previously), and its length is
    at least half of average\_length (a new condition that we introduce in
    order to ensure that a color is not associated with too few countries).
\item The chunk is already twice the average\_length (a new
    condition to ensure that we don't have too many countries associated with
    the same color).
\item The chunk's area reaches double AVG (i.e., it's getting too big).
\end{enumerate}
It is important to note that as the chunks are not necessarily evenly-balanced,
if we blindly follow the above conditions, then we might run out of elements
before we get to the last chunk.  We mitigate this issue by determining the
breaks recursively.  In particular, each time we find a break, we recursively
apply the algorithm to the rest of the elements.  The pseudocode for the
algorithm is given by Algorithm~\ref{alg:balanced-greedy}.  It is a recursive
procedure that finds the greedy breaks as described above.  It is invoked using
\Call{OptimizedGreedy}{areas, 0, $K$} and it returns the indices of
the $K$ breaks.

\subsection{Optimized Dynamic Programming Algorithm}
\label{sec-dynamic-programming-algorithm}
Recall that when we discussed equal area algorithm, we optimized: $ERROR\left(
{breaks} \right) = \frac{1}{K} \sum_{i = 1}^K {\left| {sum\left(
{chun{k_i}} \right) - AVG} \right|} $.  We now modify this expression to also
account for the lengths of the chunks as well.  Therefore, our goal is to
minimize the following formula.
$$(1 - W) \sum\limits_{i = 1}^K {\left| \dfrac{\textrm{sum}(chunk_i) - AVG}{K \cdot AVG} \right|}^2 + W \sum\limits_{i = 1}^K {\left| \dfrac{\textrm{len}(chunk_i) - \frac{N}{K}}{N} \right|}^2$$
$W$ is a user-defined constant specifying the weight to be given to the
lengths.  We use the term {\em W-score} to describe it on account of its
similarity to the concept of an {\em f-score} used in information retrieval to
vary the influence of precision and recall~\cite{Salt89}.  The left term in the
summation corresponds to the normalized average error in area, while the right
term in the summation corresponds to the normalized average error in length.
Therefore, setting $W=0$ yields the equal area algorithm, while setting $W=1$
yields the equal length (quantiles) algorithm.  On the other hand, setting
$W=0.5$ yields an intermediate measure between equal area and equal length,
which is what we are seeking.

Again, as in Section~\ref{sec-dynamic-programming-equal-area} we use dynamic
programming to optimize the new criterion.  However, the trick that we used to
find only two candidate locations for the last break will not work here, as
moving the break to the left or the right results not only in changing the sums
of the chunks, but also in changing their lengths.  To overcome this, we
consider all the locations for each break, at the cost of a higher complexity
of $O(N^2 K)$.  The pseudocode for the algorithm is given by
Algorithm~\ref{alg:dp-balanced}

\begin{algorithm}[!t]
\caption{Optimized dynamic programming algorithm for partitioning a sequence of
    $N$ numbers into $K$ chunks with roughly equal sum, while simultaneously
    balancing the lengths of the chunks.  The result is something between
    equal length (quantiles) and equal area. $W$ specifies the desired weight
    to be given to the lengths factor. The algorithm returns the breaks. Complexity: $O(N^2 K)$.}
\label{alg:dp-balanced}
\begin{algorithmic}[1]
\Statex
\Function{DPOptimized}{numbers}.
\Let{$AVG$}{sum($numbers$) / $K$}  \Comment Average chunk sum.
\Let{$AVG\_LEN$}{$N / K$}  \Comment Average chunk length.
\Let{$best\_error$}{
$<$\parbox[t]{\dimexpr 17em}{
2D array with $N$+1 rows (from 0 to $N$) for the end index $m$,
and $K$ columns (from 0 to $K$-1) for the number of breaks $b$.$>$}
}
\Let{$best\_breaks$}{
$<$\parbox[t]{\dimexpr 16em}{2D array like above, for the breaks$>$}
}

\For{$m$ in $0..N$} \Comment Fill the first column (column 0).
	\Let{$best\_error$[$m$][0]}{\parbox[t]{\dimexpr 15em}{
		$(1 - W) \left|\dfrac{\Call{PSum}{0, m} - AVG}{\Call{PSum}{0, N}}\right|^2
		+ W \left|\dfrac{m - AVG\_LEN}{N}\right|^2$}}
	\Let{$best\_breaks[m][0]$}{[]}
\EndFor

\For{$b$ in $1..K-1$}
	\For{$m$ in $0..N$}
		\Let{$min\_error$}{uninitialized}
		\Let{$best\_break$}{uninitialized}
		
		\For{$break$ in $0..m$}
			\Let{$break\_error$}{$best\_error[break][b - 1]$
				\NoNumber{\tab $+ (1 - W) \left|\dfrac{\Call{PSum}{break, m} - AVG}{\Call{PSum}{0, N}}\right|^2$}
				\NoNumber{\tab $+ W \left|\dfrac{(m - break) - AVG\_LEN}{N}\right|^2$}}
			
			\If{$break\_error < min\_error$}
				\Let{$min\_error$}{$break\_error$}
				\Let{$best\_break$}{$break$}
			\EndIf
		\EndFor
		\Let{$best\_error[m][b]$}{$min\_error$}
		\Let{$best\_breaks[m][b]$}{\parbox[t]{\dimexpr 12.2em}{
		$best\_breaks[best\_break][b\textrm{ - 1}] + [best\_break]$}}
	\EndFor
\EndFor
\State \Return $best\_breaks[N][K - 1]$
\EndFunction
\end{algorithmic}
\end{algorithm}

\begin{figure}[t!]
\centering
  \begin{subfigure}[t]{0.7\columnwidth}
    \includegraphics[width=\columnwidth]{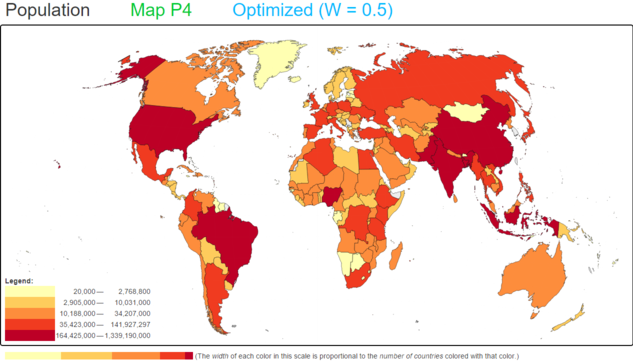}
    \caption{Population}
  \end{subfigure}

  \begin{subfigure}[t]{0.7\columnwidth}
    \includegraphics[width=\columnwidth]{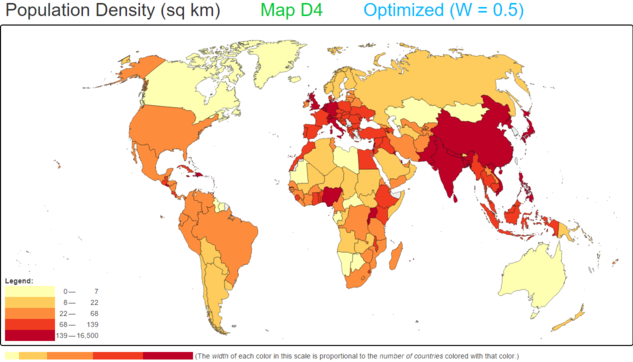}
    \caption{Population density}
  \end{subfigure}

  \begin{subfigure}[t]{0.7\columnwidth}
   \includegraphics[width=\columnwidth]{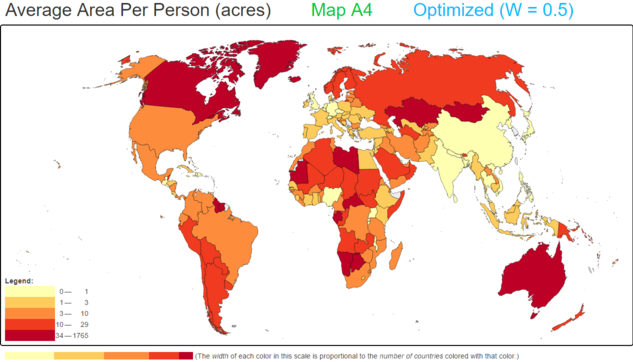}
   \caption{Area per person}
  \end{subfigure}
  \caption{Statistics mapped using the optimized algorithm with $W=0.5$}
\label{fig:opt-maps}
\end{figure}

Figure~\ref{fig:opt-maps} shows Choropleth maps using the optimal algorithm
with $W=0.5$ for the population, population density, and average area per
person using the Winkel-Tripel projection.  Here we see that varying $W$ in
this way increases (reduces) the proportion of lightly colored countries in the
population map when using the equal length (equal area) algorithm, reduces
(increases) the proportion of lightly colored countries in the population
density map when using the equal length (equal area) algorithm, and increases
(reduces) the proportion of lightly colored countries in the average area per
person map when using the equal length (equal area) algorithm.  Readers can see
this by making use of the $W$ varying the slider tool available at our online
tool\footnote{\url{www.visumaps.com/more/final\_surveys/W\_BlackBorders.html}}.
In fact, when we asked 15 arbitrary people on Amazon Mechanical Turk for their
preferred values for $w$, we received different preferred values for $W$. The
most popular answers were 0.3, 1, and 0.4 for the population, population
density, and average area per person datasets, respectively.

\section{Concluding Remarks}
\label{sec-concluding-remarks}

It is worth noting that generally-speaking, larger countries tend to have
higher population and lower population densities. Since the equal length
algorithm assigns colors so that an equal number of countries are
assigned to each color, many large countries are assigned dark colors for the
population map , and many large countries are assigned light colors for the
population density map, resulting in a fairly dark population map and a light
population density map.  Users seemed to prefer the lighter maps which led us
to believe that people prefer maps that are dominated by light colors rather
than dark colors.  This theory needs to be tested in future studies.  It is
also interesting to research whether changing the color of the borders between
regions from black to white has any effect on the results.

The feedback we received also revealed that some users prefer more variability
(contrast) between the colors of adjacent countries in order to facilitate the
easier discovery of differences. In particular, the criteria that we discussed
in this paper could also be modified to minimize the number of adjacent regions
with the same color.

We also believe that people like to be able to easily identify extremes on the
map (e.g., densest country). To address such an issue, we could devise an
algorithm that assigns the colors based on a variant of a bell curve, such that
fewer regions fall in the initial and final color ranges.  Another direction
for future work is to devise an algorithm that exploits the middle ground
between the equal area and equal length algorithms that ignore the values when
partitioning, and clustering algorithms such as the Jenks natural breaks
algorithm which focus on finding natural breaks in the values.

%
%

It is interesting to note that the equal area and equal length methods are
similar in spirit to spatial indexing methods that are differentiated on the
basis of whether they organize the underlying embedding space  (region
quadtrees) or the underlying data (point quadtrees)~\cite{Same04}. In the
former case, the areas for each color are the same while in the latter the
number of objects associated with each color are the same.  We used a
similarly-spirited analogy to compare the application of the equal area method
to magnitude data to the use of normalized data such as densities in
Section~\ref{sec-density-maps}.  In particular, we pointed out that the
equal area method applies the normalization to the embedding space (i.e., the
world map) for which the data has been collected to yield equal-sized areas,
while density measures normalize the measured quantity over an area.  Thus the
equal area method can be said to be an alternative to the argument that raw
data values should be normalized using measures such as densities before
visualizing them with a Choropleth map.

\bibliographystyle{abbrvnat}
\fontsize{8}{9.5}\selectfont
\setlength{\bibsep}{0pt plus 0.3ex}
\bibliography{hjs,equal_area_breaks}

\begin{thebibliography}{29}
\providecommand{\natexlab}[1]{#1}
\providecommand{\url}[1]{\texttt{#1}}
\expandafter\ifx\csname urlstyle\endcsname\relax
  \providecommand{\doi}[1]{doi: #1}\else
  \providecommand{\doi}{doi: \begingroup \urlstyle{rm}\Url}\fi

\bibitem[Armstrong et~al.(2003)Armstrong, Xiao, and
  Bennett]{armstrong_using_2003}
M.~P. Armstrong, N.~Xiao, and D.~A. Bennett.
\newblock Using {Genetic} {Algorithms} to {Create} {Multicriteria} {Class}
  {Intervals} for {Choropleth} {Maps}.
\newblock \emph{Annals of the Association of American Geographers}, 93\penalty0
  (3):\penalty0 595--623, Sept. 2003.

\bibitem[Brewer(1994)]{brewer_color_1994}
C.~A. Brewer.
\newblock Color {Use} {Guidelines} for {Mapping} and {Visualization}.
\newblock \emph{Modern Cartography Series}, pages 123--147, Jan. 1994.

\bibitem[Brewer and Pickle(2002)]{brewer_evaluation_2002}
C.~A. Brewer and L.~Pickle.
\newblock Evaluation of {Methods} for {Classifying} {Epidemiological} {Data} on
  {Choropleth} {Maps} in {Series}.
\newblock \emph{Annals of the Association of American Geographers}, 92\penalty0
  (4):\penalty0 662--681, Dec. 2002.

\bibitem[Carr et~al.(1992)Carr, Olsen, and White]{carr_hexagon_1992}
D.~B. Carr, A.~R. Olsen, and D.~White.
\newblock Hexagon {Mosaic} {Maps} for {Display} of {Univariate} and {Bivariate}
  {Geographical} {Data}.
\newblock \emph{Cartography and Geographic Information Systems}, 19\penalty0
  (4):\penalty0 228--236, Jan. 1992.

\bibitem[{Central Intelligence
  Agency}(2019)]{central_intelligence_agency_country_2019}
{Central Intelligence Agency}.
\newblock The world factbook 2019 --- country comparison :: {GDP} - per capita
  ({PPP}), 2019.

\bibitem[Chang and Schiewe(2018)]{chang_task_2018}
J.~Chang and J.~Schiewe.
\newblock Task-oriented {Data} {Classification} of {Choropleth} {Maps} for
  {Preserving} {Local} {Extreme} {Values}.
\newblock In \emph{Photogrammetrie - {Fernerkundung} - {Geoinformatik} -
  {Kartographie}}, pages 213--218, Mar. 2018.

\bibitem[David M.~Goldberg(2007)]{Gold07}
I.~David M.~Goldberg, J. Richard~Gott.
\newblock {Flexion and Skewness in Map Projections of the Earth}.
\newblock \emph{Cartographica}, 42\penalty0 (4):\penalty0 297--318, 2007.

\bibitem[Du et~al.(2018)Du, Ren, Zhou, Li, Tian, and Dai]{du_banded_2018}
Y.~Du, L.~Ren, Y.~Zhou, J.~Li, F.~Tian, and G.~Dai.
\newblock Banded {Choropleth} {Map}.
\newblock \emph{Personal Ubiquitous Computing}, 22\penalty0 (3):\penalty0
  503--510, June 2018.

\bibitem[Harrower and Brewer(2003)]{harrower03}
M.~Harrower and C.~A. Brewer.
\newblock Colorbrewer.org: An online tool for selecting colour schemes for
  maps.
\newblock \emph{The Cartographic Journal}, 40\penalty0 (1):\penalty0 27--37,
  2003.

\bibitem[Jenks(1977)]{Jenk77}
G.~F. Jenks.
\newblock {Optimal data classification for choropleth maps}.
\newblock Geography Department Occasional Paper No. 2~2, University of Kansas,
  Lawrence, KS, 1977.

\bibitem[Jenks and Caspall(1971)]{jenks_error_1971}
G.~F. Jenks and F.~C. Caspall.
\newblock Error on {Choroplethic} {Maps}: {Definition}, {Measurement},
  {Reduction}.
\newblock \emph{Annals of the Association of American Geographers}, 61\penalty0
  (2):\penalty0 217--244, 1971.

\bibitem[Jern et~al.(2009)Jern, Rogstadius, and Åström]{jern_treemaps_2009}
M.~Jern, J.~Rogstadius, and T.~Åström.
\newblock Treemaps and {Choropleth} {Maps} {Applied} to {Regional}
  {Hierarchical} {Statistical} {Data}.
\newblock In \emph{Proceedings of the 13th International Conference on
  Information Visualisation, {IV}}, pages 403--410, Bacelona, Spain, July 2009.

\bibitem[Jiang(2013)]{Jian13}
B.~Jiang.
\newblock {Head/tail breaks: A new classification scheme for data with a
  heavy-tailed distribution}.
\newblock \emph{The Professional Geographer}, 65\penalty0 (3):\penalty0
  482--494, 2013.

\bibitem[{Johnson} and {Shneiderman}(1991)]{johnson_treemaps_1991}
B.~{Johnson} and B.~{Shneiderman}.
\newblock Tree-maps: a space-filling approach to the visualization of
  hierarchical information structures.
\newblock In \emph{Proceedings of IEEE Visualization '91}, pages 284--291, San
  Diego, CA, Oct. 1991.

\bibitem[{Lima} et~al.(2019){Lima}, {Leal}, d.~S.~{Brito}, {Resque dos Santos},
  and {Meiguins}]{lima_chorolibre_2019}
R.~{Lima}, M.~S.~B. {Leal}, Y.~P. d.~S.~{Brito}, C.~G. {Resque dos Santos}, and
  B.~S. {Meiguins}.
\newblock {ChoroLibre}: Supporting georeferenced demographic information
  visualization through hierarchical choropleth maps.
\newblock In \emph{Proceedings of 23rd International Conference in Information
  Visualization – Part II}, pages 56--61, Adelaide, Australia, July 2019.

\bibitem[Lloyd and Steinke(1976)]{lloyd_decision_1976}
R.~Lloyd and T.~Steinke.
\newblock The {Decision} making {Process} for {Judging} the {Similarity} of
  {Choropleth} {Maps}.
\newblock \emph{The American Cartographer}, 3\penalty0 (2):\penalty0 177--184,
  1976.

\bibitem[Lloyd and Steinke(1977)]{lloyd_visual_1977}
R.~Lloyd and T.~Steinke.
\newblock Visual and {Statistical} {Comparison} of {Choropleth} {Maps}.
\newblock \emph{Annals of the Association of American Geographers}, 67\penalty0
  (3):\penalty0 429--436, 1977.

\bibitem[McNabb et~al.(2018)McNabb, Laramee, and Fry]{mcnabb_dynamic_2018}
L.~McNabb, R.~S. Laramee, and R.~Fry.
\newblock Dynamic {Choropleth} {Maps} – {Using} {Amalgamation} to {Increase}
  {Area} {Perceivability}.
\newblock In \emph{Proceedings of the 22nd International Conference Information
  Visualisation, {IV}}, pages 284--293, Fisciano, Italy, July 2018.

\bibitem[Murray and Shyy(2000)]{murray_integrating_2000}
A.~T. Murray and T.-K. Shyy.
\newblock Integrating attribute and space characteristics in choropleth display
  and spatial data mining.
\newblock \emph{International Journal of Geographical Information Science},
  14\penalty0 (7):\penalty0 649--667, Oct. 2000.

\bibitem[Robinson et~al.(1984)Robinson, Sale, Morrison, and
  Muehrcke]{robinson_elements_1984}
A.~H. Robinson, R.~D. Sale, J.~L. Morrison, and P.~C. Muehrcke.
\newblock \emph{Elements of cartography}.
\newblock Wiley, New York, 5th ed. edition, 1984.

\bibitem[Salton(1989)]{Salt89}
G.~Salton.
\newblock \emph{Automatic Text Processing: The Transformation Analysis and
  Retrieval of Information by Computer}.
\newblock Addison-Wesley, Reading, MA, 1989.

\bibitem[Samet(2006)]{Same04}
H.~Samet.
\newblock \emph{Foundations of Multidimensional and Metric Data Structures}.
\newblock Morgan-Kaufmann, San Francisco, 2006.
\newblock (Translated to Chinese ISBN 978-7-302-22784-7).

\bibitem[Schiewe(2019)]{schiewe_empirical_2019}
J.~Schiewe.
\newblock Empirical {Studies} on the {Visual} {Perception} of {Spatial}
  {Patterns} in {Choropleth} {Maps}.
\newblock \emph{Journal of Cartography and Geographic Information}, 69\penalty0
  (3):\penalty0 217--228, Sept. 2019.

\bibitem[Slocum et~al.(2009)Slocum, McMaster, Kessler, and
  Howard]{slocum_thematic_2009}
T.~A. Slocum, R.~B. McMaster, F.~C. Kessler, and H.~H. Howard.
\newblock \emph{Thematic {Cartography} and {Geovisualization}}.
\newblock Prentice {Hall} series in geographic information science. Pearson
  Prentice Hall, Upper Saddle River, NJ, 3rd ed. edition, 2009.

\bibitem[Speckmann and Verbeek(2010)]{speckmann_necklace_2010}
A.~Speckmann and K.~Verbeek.
\newblock Necklace {Maps}.
\newblock \emph{IEEE Transactions on Visualization and Computer Graphics},
  16\penalty0 (6):\penalty0 881--889, Nov. 2010.

\bibitem[Steinke and Lloyd(1983)]{steinke_judging_1983}
T.~Steinke and R.~Lloyd.
\newblock Judging {The} {Similarity} of {Choropleth} {Map} {Images}.
\newblock \emph{Cartographica: The International Journal for Geographic
  Information and Geovisualization}, 20\penalty0 (4):\penalty0 35--42, 1983.

\bibitem[Tobler(2004)]{Cartogram}
W.~Tobler.
\newblock Thirty-five years of computer cartograms.
\newblock \emph{Annals of the Association of American Geographers},
  94:\penalty0 58--73, 2004.

\bibitem[Xiao and Armstrong(2006)]{xiao_choroware_2006}
N.~Xiao and M.~P. Armstrong.
\newblock {ChoroWare}: {A} {Software} {Toolkit} for {Choropleth} {Map}
  {Classification}.
\newblock \emph{Geographical Analysis}, 38\penalty0 (1):\penalty0 102--121,
  2006.

\bibitem[{Zhang} and {Maciejewski}(2017)]{zhang_quantifying_2017}
Y.~{Zhang} and R.~{Maciejewski}.
\newblock Quantifying the visual impact of classification boundaries in
  choropleth maps.
\newblock \emph{IEEE Transactions on Visualization and Computer Graphics},
  23\penalty0 (1):\penalty0 371--380, Aug. 2017.

\end{thebibliography}

\end{document}